\pdfoutput=1

\documentclass[aps,twocolumn,amsmath,amssymb,preprintnumbers,floatfix,prb,superscriptaddress,longbibliography]{revtex4-2}

\usepackage[utf8]{inputenc}
\usepackage{newtxtext}
\usepackage[upint]{newtxmath}
\usepackage{microtype}
\usepackage{textcomp}
\usepackage{eucal}
\usepackage{bm}
\usepackage{siunitx}
\usepackage{comment}
\usepackage{lipsum}
\usepackage{mathtools}
\usepackage{tikz}
\usetikzlibrary{calc,decorations.pathmorphing}
\usepackage{enumerate}
\usepackage{amsfonts}
\usepackage{amsmath}
\usepackage{amssymb}
\usepackage{color}
\usepackage{soul}
\usepackage{layouts}

\usepackage{graphicx}

\usepackage[colorlinks,allcolors=blue]{hyperref}
\usepackage[capitalize]{cleveref}

\definecolor{DarkRed}{rgb}{0.65,0,0}%
\definecolor{Green}{rgb}{0,0.3,0.3}
\definecolor{Purple}{rgb}{0.3,0,0.65}
\definecolor{Red}{rgb}{1,0,0}
\definecolor{Blue}{rgb}{0,0,0.85}
\definecolor{Magenta}{rgb}{1,0,1}

\newcommand{\ca}[2][]{c_{#2}^{\vphantom{\dagger}#1}} 
\newcommand{\cc}[2][]{c_{#2}^{{\dagger}#1}}          

\newcommand{\eg}{\textit{e.g. }}

\newcommand{\be}{\begin{equation}}
\newcommand{\ee}{\end{equation}}

\newcommand{\prlsection}[1]{\textit{#1}.\kern0.05em---\kern0.05em\ignorespaces}

\begin{document}

\title{Critical temperature of triplet superconductor-ferromagnet bilayers as a probe for pairing symmetry}
\author{Erik Wegner Hodt}
\affiliation{Center for Quantum Spintronics, Department of Physics, Norwegian \\ University of Science and Technology, NO-7491 Trondheim, Norway}

\author{Carla Cirillo}
\affiliation{CNR-SPIN, c/o Universit\`{a} degli Studi di Salerno - Via Giovanni Paolo II,
132 - I-84084 - Fisciano (Sa), Italy}

\author{Angelo Di Bernardo}
\affiliation{Dipartimento di Fisica “E.R. Caianiello”,
Universit\`{a} degli Studi di Salerno - Via Giovanni Paolo II,
132 - I-84084 - Fisciano (Sa), Italy}

\author{Carmine Attanasio}
\affiliation{Dipartimento di Fisica “E.R. Caianiello”,
Universit\`{a} degli Studi di Salerno - Via Giovanni Paolo II,
132 - I-84084 - Fisciano (Sa), Italy}

\author{Jacob Linder}
\affiliation{Center for Quantum Spintronics, Department of Physics, Norwegian \\ University of Science and Technology, NO-7491 Trondheim, Norway}

\date{\today}
\begin{abstract}
Identifying superconducting materials with spin-polarized Cooper pairs is an important objective both for exploration of new fundamental physics and for cryogenic applications in spintronics and quantum sensing. We here compute the critical temperature $T_c$ of the superconducting transition in a bilayer comprised of a superconductor with an intrinsic spin-triplet order parameter and a ferromagnet. We determine how $T_c$ varies both with the thickness of the ferromagnet and its magnetization direction. We show that both the orbital and spin part of the triplet superconducting order parameter leave clear signatures in $T_c$ which do not appear in a bilayer of a conventional $s$-wave superconductor and a ferromagnet. In particular, the dependence of $T_c$ on these variables changes depending on whether or not the superconducting order parameter features Andreev bound-states and also changes qualitatively when the magnetization is rotated in the plane of the ferromagnetic film.  Measurements of $T_c$ in such bilayers are therefore useful to identify the pairing symmetry of intrinsic triplet superconductors.
\end{abstract}

\maketitle
\section{Introduction}

Merging materials with fundamentally different properties can give rise to new quantum physics at the interface which is interesting both from a fundamental perspective and in terms of possible applications. A prime example of this is superconducting and magnetic materials, a combination which has been studied thoroughly in particular during the last decades \cite{buzdin_rmp_05, bergeret_rmp_05}.

The most fundamental property of superconductors is the Meissner effect, leading to expulsion of magnetic fields. Notwithstanding, it is possible for magnetic order and superconductivity to coexist in hybrid structures. The Cooper pairs in superconductors can align their spins to endure the spin-polarized environment provided by ferromagnets, opening the possibility to have dissipationless transport not only of charge, but also of spin. Spin-polarized superconductivity, known as triplet pairing, is well-established in heterostructures comprised of superconductors and ferromagnets \cite{bergeret_rmp_05, bergeret_prl_01, keizer_nature_06, khaire_prl_10, robinson_science_10, Eschrig2015a, linder_nphys_15}. Intrinsic triplet superconducting materials are, however, scarce. It has been experimentally established that uranium-based ferromagnetic superconductors such as UGe$_2$ and URhGe feature triplet pairing \cite{Saxena2000,Aoki2001, Aoki2011, Huy_prl2007}, but besides these evidence for intrisic triplet pairing in superconducting materials is mostly inconclusive. 

Triplet pairing can be put on firm experimental ground by performing measurements of multiple physical observables if all such measurements indicate the presence of spin-polarized superconductivity. It is therefore important to establish how triplet pairing is manifested in various experimentally accessible quantities. The critical temperature $T_c$ at which a material enters its superconducting phase is both a fundamental and one of the most accessible experimental quantities available. 

Motivated by this, we here report a theoretical study of how $T_c$ in a triplet superconductor is altered when placing the superconductor in contact with a ferromagnet. Such a setup, with a conventional s-wave superconductor described by Bardeen-Cooper-Schrieffer (BCS) theory rather than a triplet superconductor, has been experimentally demonstrated to yield clear evidence of proximity-induced triplet pairing \cite{cirillo_prb_05, zdravkov_prl_06, leksin_prl_12, gu_apl_14}. Here, we instead investigate how $T_c$ of intrinsically present triplet Cooper pairs changes when introduced to a ferromagnetic environment. We find that measurements of $T_c$  
 provides clear information about the nature of the superconducting triplet state. Specifically, the behavior of $T_c$ in the presence of a ferromagnet reveals information about both the orbital part and the spin part of the triplet pairing state. We therefore argue that $T_c$ measurements of candidate triplet superconducting materials placed in contact with conventional ferromagnets are a useful tool to infer information about the properties of the superconducting state and obtain evidence supporting the spin-triplet nature of the order parameter.

\begin{figure}[h!]
    \centering
    \includegraphics{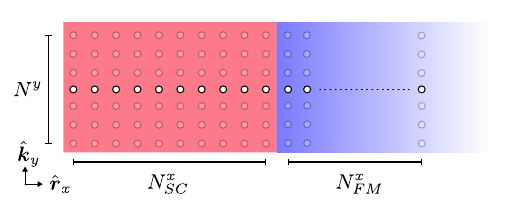}
    \caption{The bilayer geometry consisting of a p-wave triplet superconductor (red) adjacent to a ferromagnetic layer (blue) of varying thickness. The system is assumed periodic in the \textit{y}-direction parallel to the interface and of finite size in the \textit{x}-direction normal to the interface. It is thus suitable to describe the system with real-space coordinates $r_x$ in \textit{x}  and momentum-space coordinates $k_y$ in \textit{y}. $N^y$ denotes the number of momentum modes in \textit{y} while $N_{SC}^x$ and $N_{FM}^x$  denote the number of \textit{x}-direction lattice sites in the SC and FM, respectively. }
    \label{fig: system}
\end{figure}
\section{Model}
We consider a tight binding Hamiltonian on a square lattice bilayer structure consisting of a superconductor (SC) and an adjacent ferromagnet (FM). In the SC, we consider nearest neighbour attractive electron-electron interaction,
\begin{multline}
    H_\text{SC}=-\sum_{\langle i,j\rangle, \sigma}t_{ij} \cc[]{i,\sigma}\ca[]{j,\sigma}\\  - \sum_{i,\sigma}\mu \cc[]{i,\sigma}\ca[]{i,\sigma}  +  \sum_{\langle i, j \rangle, \sigma} V_{ij} n_{i,\sigma}n_{j,\sigma}
\end{multline}
where $\cc[]{i,\sigma}$ ($\ca[]{i,\sigma}$) creates (annihilates) an electron with spin $\sigma$ on site \textit{i}, $\langle i,j\rangle$ denotes all nearest neighbor site pairs, and $\mu$ is the chemical potential. We shall assume the hopping to be isotropic in the following, $t_{ij}=t>0$.  $V_{ij}\leq 0$ is the coupling strength of an effective, attractive electron-electron interaction in the system, On account of being a nearest-neighbor interaction, it can give rise to \textit{p}-wave superconductivity.  

Adjacent to the superconductor, we consider a ferromagnet modelled by the following Hamiltonian 
\begin{equation}
\begin{split}
    H_\text{FM} &=-\sum_{\langle i,j\rangle, \sigma}t  \cc[]{i,\sigma}\ca[]{j,\sigma} - \sum_{i,\sigma}\mu  \cc[]{i,\sigma}\ca[]{i,\sigma} \\&\qquad\qquad- m\sum_{i,\sigma,\sigma'}\hat{\boldsymbol{n}}\cdot\boldsymbol{\sigma}\cc[]{i,\sigma}\ca[]{i,\sigma'}
\end{split},
\end{equation}
where \textit{m} is the strength of the FM exchange field, $\hat{\boldsymbol{n}}$ is a unit vector denoting the FM magnetization direction, and $\boldsymbol{\sigma}$ is the vector of Pauli matrices. The SC and FM regions are connected by electron hopping, enabling a proximity effect between the materials, but the attractive electron-electron pairing strength $V_{ij}$ is zero in the FM while the strength of the FM exchange field \textit{m} is zero in the SC. We will for simplicity only consider the case where the chemical potential is constant across the heterostructure.

\subsection{Constructing the Hamiltonian matrix}

We shall assume the system to be periodic in \textit{y}, the direction parallel to the interface. In order to take advantage of this periodicity, we introduce momentum-space operators  
in the \textit{y}-direction as indicated by the system geometry presented in Fig. \ref{fig: system}. Given the Fourier-transformed operators, 
\begin{equation}
\begin{split}
    \ca[]{(i_x,i_y),\sigma}&=\frac{1}{\sqrt{N^y}}\sum_{k_y}\ca[]{i_x,k_y,\sigma}e^{ik_yr_y} \\ \cc[]{(i_x,i_y),\sigma}&=\frac{1}{\sqrt{N^y}}\sum_{k_y}\cc[]{i_x,k_y,\sigma}e^{-ik_yr_y},
    \end{split}
\end{equation}
the Hamiltonian, now block-diagonal in $k_y$ modes, can be written as 
\begin{align}
    H=E_0 + \frac{1}{2}\sum_{k_y}B_{k_y}^\dagger H_{k_y}B_{ k_y}^{\vphantom{\dagger}} \label{eqn: general hamiltonian}
\end{align}
where the operator-independent energy $E_0$ gives a contribution to the total system energy which is not relevant for this paper.

Each $k_y$ block in the total Hamiltonian matrix (Eq. (\ref{eqn: general hamiltonian})) is now a $4N_{SC}^x N_{FM}^x \times 4N_{SC}^x N_{FM}^x$ matrix given by
\begin{equation}
    B_{k_y}^\dagger H_{k_y} B_{ k_y}^{\vphantom{\dagger}} = \sum_{\langle i_x, j_x \rangle} \big( W_{i_x}^{k_y}\big)^\dagger h_{i_x, j_x}^{k_y}W_{j_x}^{k_y}
\end{equation}
where the sum over $i_x, j_x$ runs over the 1D string of lattice sites in the SC and FM. We shall from now on omit subscripts on real-space and momentum indices, taking a real space coordinate $i$ to implicitly mean a site in the \textit{x} direction and vice versa for momentum indices. The basis vectors $W_{i}^{k}$ are given by 
\begin{equation}
    W_{i}^{k} = \begin{pmatrix}
        \ca[]{i, k, \uparrow} & \ca[]{i, k, \downarrow} & \cc[]{i, -k, \uparrow} & \cc[]{i, -k, \downarrow}
    \end{pmatrix}^T
\end{equation}

When $i, j$ are in the SC, the matrix elements $h_{i, j}^{k}$ are defined for $i = j$ as (see appx. \ref{App: p-wave} for details)
\begin{multline}
    h_{i, i}^{k,\text{SC}} = \\  \begin{bmatrix}
        \varepsilon_{k_y} + \mu  & 0 & -2\Delta_{i, k}^{\uparrow, \hat{y}} & 0 \\
        0 & \varepsilon_{k} + \mu & 0 & -2\Delta_{i, k}^{\downarrow, \hat{y}} \\ 
        -2(\Delta_{i, k}^{\uparrow, \hat{y}})^\dagger & 0 & -(\varepsilon_{k} + \mu) & 0 \\
        0 & -2 (\Delta_{i, k}^{\downarrow, \hat{y}})^\dagger & 0 & -(\varepsilon_{k} + \mu)
    \end{bmatrix}
\end{multline}
and for $j=i\pm \hat{x}$,
\begin{multline}
h_{i, i\pm\hat{x}}^{k, \text{SC}} =\\
     \begin{bmatrix}
        -t & 0 & -2\Delta_{i}^{\uparrow, \pm\hat{x}} & 0 \\
        0 & -t & 0 & -2\Delta_{i}^{\downarrow, \pm\hat{x}} \\ 
        2(\Delta_{i}^{\uparrow, \pm\hat{x}})^\dagger & 0 & t & 0 \\
        0 & 2 (\Delta_{i}^{\downarrow, \pm\hat{x}})^\dagger & 0 & t
    \end{bmatrix}
\end{multline}
Here we have defined the order parameters (also known as gaps in the context of superconductivity)
\begin{align}
    \Delta_{i}^{\sigma,\pm\hat{x}}&=\frac{V_x}{N^y}\sum_{k'} F_{i,k'}^{\sigma, \pm\hat{x}} \\
    \Delta_{i, k}^{\sigma, \hat{y}}&= \frac{V_y}{N^y}\sum_{k'}2\cos(k'-k) F_{i,k'}^{\sigma, \hat{y}}\label{eqn: gaps}
\end{align} 
the pairing amplitudes 
\begin{align}
F_{i,k'}^{\sigma,\pm\hat{x}}&=\big\langle \ca[]{i, k', \sigma}\ca[]{i\pm\hat{x}, -k', \sigma}\big\rangle \\[10pt] F_{i,k'}^{\sigma,\hat{y}}&=\langle \ca[]{i, k', \sigma}\ca[]{i, -k', \sigma}\big\rangle,
    \end{align} 
and the \textit{y}-direction dispersion
\begin{equation}
    \varepsilon_{k}=-2t\cos{k}-\mu
\end{equation}
The matrix elements for the FM is given similarly for $i=j$ in the FM as
\begin{multline}
    h_{i, i}^\text{FM} =\\  \begin{bmatrix}
        \varepsilon_{k} +h^{\uparrow\uparrow} & h^{\uparrow\downarrow} & 0 & 0 \\
        h^{\downarrow\uparrow} & \varepsilon_{k}  +h^{\downarrow\downarrow} & 0 & 0 \\ 
        0 & 0 & -(\varepsilon_{k} + h^{\uparrow\uparrow}) & -(h^{\uparrow\downarrow})^* \\
        0 & 0 & -(h^{\downarrow\uparrow})^* & -(\varepsilon_{k} +h^{\downarrow\downarrow})
    \end{bmatrix}
\end{multline}
\\
where we have defined the short-hand matrix $h=m\hat{\boldsymbol{n}}\cdot\boldsymbol{\sigma}$. For $j=i\pm \hat{x}$, we have
\begin{equation}
    h_{i, i\pm\hat{x}}^\text{FM} = \begin{bmatrix}
        -t & 0 & 0 & 0 \\
        0 & -t & 0 & 0\\ 
        0 & 0 & t & 0 \\
        0 & 0 & 0 & t
    \end{bmatrix}
\end{equation}

\subsection{Diagonalizing the Hamiltonian}

As the Hamiltonian is diagonal in $k$, we can diagonalize the $k$ blocks of the Hamiltonian individually, allowing us to access significantly larger system sizes than in the case of real-space coordinates in two dimensions. Starting from Eq. (\ref{eqn: general hamiltonian}), it follows that we can write the matrix block $H_{k}^{\vphantom{\dagger}}=U_{k}^{\vphantom{\dagger}}\Lambda_{k}^{\vphantom{\dagger}}U_{k}^\dagger$ where the eigenvectors of $H_{k}^{\vphantom{\dagger}}$,
\begin{equation}
\begin{gathered}
    \Phi_{n,k} = \begin{bmatrix}\varphi_{n,1,k} & ... & \varphi_{n,i,k} & ... & \varphi_{n,N,k} \end{bmatrix}^T \\[10pt]
    \varphi_{n,i,k} = \begin{bmatrix}
        u_{n,i,k} & v_{n,i,k} & x_{n,i,k} & w_{n,i,k}
    \end{bmatrix}
\end{gathered}
\end{equation}
are the columns of $U_{k}^{\vphantom{\dagger}}$,
\begin{equation}
    U_{k}^{\vphantom{\dagger}} = \begin{bmatrix}
        \Phi_1 & \Phi_2 & ... & \Phi_{4N}
    \end{bmatrix}
\end{equation}
\\
and where $N=N_{SC}^x N_{FM}^x$. Using the unitary matrices $U_{k}^{\vphantom{\dagger}}$, we may now diagonalize the $H_{k}^{\vphantom{\dagger}}$ block, 
\begin{align}
    H&=E_0 + \frac{1}{2}\sum_{k}\Gamma_{k}^\dagger \Lambda_{k}^{\vphantom{\dagger}}\Gamma_{ k}^{\vphantom{\dagger}} \\
    &=E_0 + \frac{1}{2}\sum_{k}E_{n,k}\gamma_{n,k}^\dagger \gamma_{n,k}^{\vphantom{\dagger}}\label{eqn: diagonalized Hamiltonian} 
\end{align}
where we have introduced the new Bogoliubov operators 
\begin{align}
\Gamma_{k} = U_{k}^\dagger B_{k} \Rightarrow B_{k}=U_{k}\Gamma_{k} \label{eqn: quasiparticle operators}
\end{align}
and where $\Gamma_{k}=\begin{bmatrix}\gamma_{1,k} & \gamma_{2,k} & ... & \gamma_{4N, k}\end{bmatrix}^T$. Using Eq. (\ref{eqn: quasiparticle operators}), we can express the original fermion operators in terms of the new Bogoliubov operators,
\begin{equation}
\begin{split}
\ca[]{i,k, \uparrow} &= \sum_{n=1}^{4N}u_{n,i,k}^{\vphantom{\dagger}}\gamma_{n, k}^{\vphantom{\dagger}}, \quad \ca[]{i,k, \downarrow} = \sum_{n=1}^{4N}v_{n,i, k}^{\vphantom{\dagger}}\gamma_{n,k}^{\vphantom{\dagger}}  \\
\cc[]{i,-k, \uparrow} &= \sum_{n=1}^{4N} w_{n,i, k}^{\vphantom{\dagger}} \gamma_{n,k}^{\vphantom{\dagger}}, \quad
\cc[]{i,-k, \downarrow} = \sum_{n=1}^{4N}x_{n,i,k}^{\vphantom{\dagger}} \gamma_{n,k}^{\vphantom{\dagger}} 
\end{split} \label{eqn: new operators}
\end{equation}

There are $4N$  operators in the Hamiltonian, but only $2N$ are independent. To regain the correct number of degrees of freedom, we proceed by using that positive and negative momentum operators and eigenvalues are related by the following relations, 
\begin{equation}
    \gamma_{n,k}^{\vphantom{\dagger}} = \gamma_{n,-k}^\dagger, \qquad E_{n,k} = -E_{n,-k}
\end{equation} \\
In order to incorporate this dependence between operators, we split the sum over momentum modes in the diagonalized Hamiltonian (Eq. (\ref{eqn: diagonalized Hamiltonian}))  into three parts, $k>0$, $k=0$ and $k<0$, and rewrite it as a sum over only $k=0$ and $k>0$ modes, making use of the above relations. The Hamiltonian expressed in terms of independent operators $\gamma$/$\gamma^\dagger$ then becomes
\begin{multline}
    H=E_0 + \frac{1}{2}\sum_{n,k>0}E_{n,k} + \frac{1}{2}\sum_{E_{n,0}\geq0}E_{n,0} \\
    + \sum_{n,k>0}E_{n,k}\gamma_{n,k}^\dagger \gamma_{n,k}^{\vphantom{\dagger}}+\sum_{E_{n,0}\geq 0}E_{n,0} \gamma_{n,0}^\dagger \gamma_{n,0}^{\vphantom{\dagger}}
\end{multline}
The self-consistent expressions for the gaps are now obtained by using the transformed operators from Eq. (\ref{eqn: new operators}) in the pairing amplitude expressions from Eq. (\ref{eqn: gaps}),
\begin{widetext} 
\begin{equation}\begin{split}
F_{i, k}^{\uparrow,\pm\hat{x}}&=
\frac{1}{N^y}\bigg(\sum_{k'>0,n}\bigg[(w_{n,i,k'}^* u_{n,i\pm\hat{x}, k'}^{\vphantom{*}}-u_{n,i, k'}^{\vphantom{*}}w_{n,i\pm\hat{x}, k'}^*)f(E_{n,k'}) +u_{n,i, k'}^{\vphantom{*}}w_{n,i\pm\hat{x}, k'}^* \bigg]
\\ &\qquad\qquad\qquad\qquad\qquad\qquad +\sum_{E_{n,0}\geq 0}\bigg[ (w_{n,i,0}^* u_{n,i\pm\hat{x}, 0}^{\vphantom{*}}-u_{n,i, 0}^{\vphantom{*}}w_{n,i\pm\hat{x}, 0}^*)f(E_{n,0}) +u_{n,i, 0}w_{n,i\pm\hat{x}, 0}^*\bigg]\bigg)
\end{split}\end{equation}
\begin{equation}\begin{split}
F_{i, k}^{\downarrow,\pm\hat{x}}&=
\frac{1}{N^y}\bigg(\sum_{k'>0,n}\bigg[(x_{n,i,k'}^* v_{n,i\pm\hat{x}, k'}^{\vphantom{*}}-v_{n,i, k'}^{\vphantom{*}}x_{n,i\pm\hat{x}, k'}^*)f(E_{n,k'}) +v_{n,i, k'}x_{n,i\pm\hat{x}, k'}^* \bigg]
\\ &\qquad\qquad\qquad\qquad\qquad\qquad +\sum_{E_{n,0}\geq 0}\bigg[(x_{n,i,0}^* v_{n,i\pm\hat{x}, 0}^{\vphantom{*}}-v_{n,i, 0}^{\vphantom{*}}x_{n,i\pm\hat{x}, 0}^*)f(E_{n,0}) +v_{n,i, 0}x_{n,i\pm\hat{x}, 0}^*\bigg]\bigg)
\end{split}\end{equation}
\begin{equation}\begin{split}
F_{i, k}^{\uparrow, \hat{y}}&= \frac{2}{N^y}\bigg(\sum_{k'>0}\bigg[\big(\cos(-k'-k)w_{n,i,k'}^* u_{n,i, k'}^{\vphantom{*}}-\cos(k'-k)u_{n,i, k'}^{\vphantom{*}}w_{n,i, k'}^*\big)f(E_{n,k'})\\&\qquad\qquad\qquad\qquad\qquad\qquad+ \cos(k'-k)u_{n,i, k'}^{\vphantom{\dagger}}w_{n,i, k'}^* \bigg]
+\sum_{E_{n,0}\geq0}\cos(k)u_{n,i, 0}^{\vphantom{\dagger}}w_{n,i, 0}^*\bigg)
\end{split}\end{equation}
\begin{equation}\begin{split}
F_{i, k}^{\downarrow, \hat{y}}&= \frac{2}{N^y}\bigg(\sum_{k'>0}\bigg[\big(\cos(-k'-k)x_{n,i,k'}^* v_{n,i, k'}^{\vphantom{*}}-\cos(k'-k)v_{n,i, k'}^{\vphantom{*}}x_{n,i, k'}^*\big)f(E_{n,k'})\\&\qquad\qquad\qquad\qquad\qquad\qquad+ \cos(k'-k)v_{n,i, k'}^{\vphantom{\dagger}}x_{n,i, k'}^* \bigg]
+\sum_{E_{n,0}\geq0}\cos(k)v_{n,i, 0}^{\vphantom{\dagger}}x_{n,i, 0}^*\bigg)
\end{split}\end{equation}
\end{widetext}

In order to characterize the effect of the adjacent FM on the superconductivity in the SC, we will consider the local density of states (LDOS) $N_{i}(E)$ on the lattice. To obtain an expression for the LDOS, we note that the number of charges at all sites with lattice index $i=i_x$ in the $x$-direction can be obtained by summing over all corresponding sites in the \textit{y}-direction. In effect, we will consider
\begin{equation}
\rho_{i} = \sum_{i_y,\sigma}\rho_{i,i_y}=\sum_{\sigma}\sum_k \langle \cc[]{i,k,\sigma}\ca[]{i,k,\sigma} \rangle
\end{equation}
where we have transformed to the momentum-description in the \textit{y} direction. To obtain an analytic expression for $N_i(E)$, we use that the charge number at site \textit{i} as given above is defined by an energy integral of $N_i(E)$ weighted by the Fermi-Dirac distribution function $f(E)$,
\begin{multline} 
\rho_{i}= \sum_{k}\sum_{\sigma}\langle\cc[]{i, k, \sigma}\ca[]{i, k, \sigma}\rangle = \int_{-\infty}^{\infty}dE N_{i}(E)f(E)
\end{multline}
Comparing the two formulations for the charge number $\rho_{i}$ and using the relation between the positive and negative momentum operators discussed above, the LDOS may be written as 
\begin{multline}
    N_{i}(E) = \sum_{k>0,n}\big[\big(|u_{n,i, k}|^2 +|v_{n,i,k}|^2\big)\delta(E-E_{n,k}) \\+\big(|w_{n,i,k}|^2 + |x_{n,i,k}| \big)\delta(E+E_{n,k})\big] \\+ \sum_{E_{n,0}\geq 0, n}\big[\big(|u_{n,i, 0}|^2 +|v_{n,i,0}|^2\big)\delta(E-E_{n,0}) \\+\big(|w_{n,i,0}|^2 + |x_{n,i,0}| \big)\delta(E+E_{n,0})\big]
\end{multline}

\begin{figure*}[htb]
    \centering
    \includegraphics[scale=0.99]{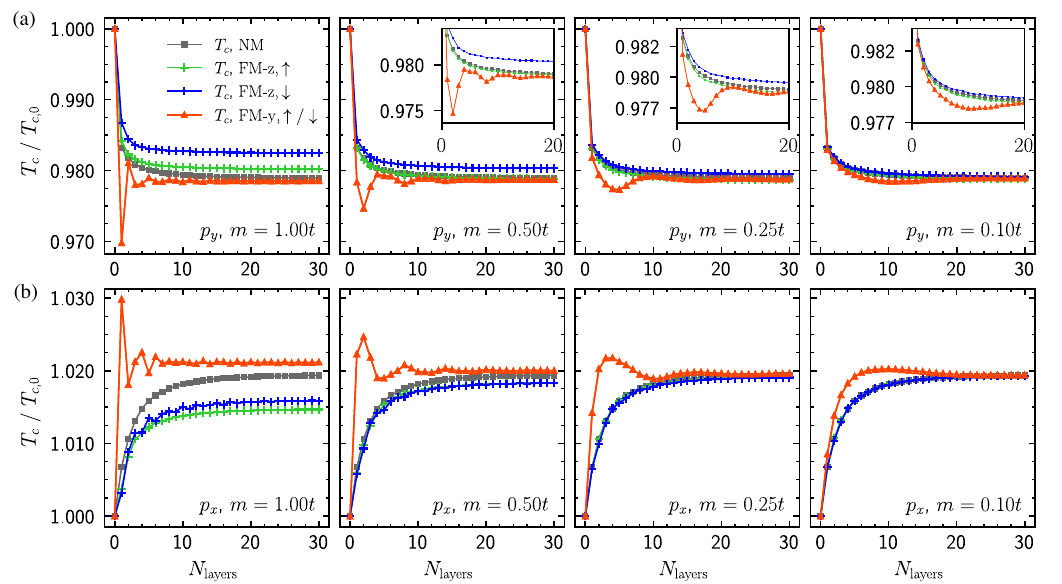}
    \caption{
    Normalized critical temperature $T_c$\, 
    is shown as function of number of NM/FM layers for $p_x$- and $p_y$-type pairing for a pairing strength $V_x=V_y=-0.85t$. The system considered consists of $N_{SC}^x=30$ sites in the superconductor and $N^y=121$ momentum modes in the \textit{y}-direction. Using a simple estimate for the superconducting coherence length $\xi$ (see Fig. \ref{fig: coherence length}), the size of the superconducting region corresponds to 2-3 $\xi$ along the direction normal to the interface. (a) For $p_y$ pairing, we observe $T_c$ exponentially decaying with added NM layers. For FM, we observe an oscillatory $T_c$ behaviour for FM polarized in \textit{y} while the \textit{z} polarized FM shows a NM-like decay, but which converges to a shifted, spin-dependent critical temperature.   (b) For $p_x$ pairing, we predict an increase in $T_c$ for both FM and NM layers. For FM polarized in \textit{y}, $T_c$ is observed to oscillate as function of layer number with an oscillation frequency decreasing with exchange field strength. The FM polarized in \textit{z} shows a near monotonous decay, resembling the normal metal case, but with a similarly shifted asymptotic $T_c$ as for $p_y$ pairing.  The general increase in $T_c$ with the addition of layers is generally attributed to a decrease in the suppressive effect of Andreev bound states on the gap in the SC. The insets in the top row show a zoom-in of each panel.}
    \label{fig: t_c vs thickness}
\end{figure*}
\begin{figure}
    \centering
    \includegraphics{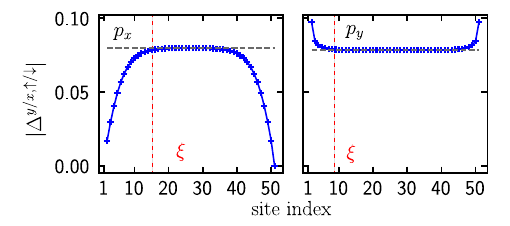}
    \caption{The spatial dependence of the SC gap is shown for a bare SC with $N_{SC}^x=50$ and $N^y=121$ for $p_x$ and $p_y$ pairing. By a visual inspection of the gap magnitude, we can estimate the superconducting coherence length $\xi$ to be around 15 sites for $p_x$ pairing and 10 sites for $p_y$ pairing, giving a coarse length scale at which we expect proximity effects to be significant.}
    \label{fig: coherence length}
\end{figure}
\subsection{Binary search algorithm}
In order to obtain an estimate for $T_c$ of the SC/FM bilayer we utilize a binary search algorithm \cite{ouassou2015density, Johnsen2020:PRL}. The usual self-consistency scheme would entail obtaining a self-consistent gap for all temperatures within a range $[T_{min}, T_{max}]$ and estimating $T_c$ as the temperature at which the gap crosses above a predetermined threshold. If we for instance denote a self-consistent solution to either of the gaps from Eq. \ref{eqn: gaps} obtained at $T=0$ as $\Delta_0$, then the $T_c$ threshold could be the temperature at which the gap falls below $\Delta_0/1000$.  An accurate estimate for $T_c$ is computationally expensive with this approach. Instead, we perform a binary search algorithm where only a handful of self-consistent iterations are performed for each parameter configuration. 

To obtain an estimate for $T_c$, we proceed in the following way: one first obtains the fully self-consistent $T=0$ gap $\Delta_0$ for a temperature well below $T_c$, denoted $T_{min}$. Based on the system parameters, one can then select a $T_{max}$ well above $T_c$ and discretize the range $[T_{min}, T_{max}]$ into a given number of discrete temperature steps, defining a range in which $T_c$ is expected to be. A binary search is now performed on this temperature range, using a small fraction $\Delta_0/1000$ of the self-consistent solution as the initial value of the gap. After $n$ iterations ($n=10$ in calculations presented in this paper), a check is performed in order to determine whether the gap has increased above $\Delta_0/1000$, implying that we are below $T_c$, or decreased below $\Delta_0$, implying that we are above $T_c$. The temperature value at which the binary search algorithm converges is our estimate for the critical temperature. As probe for the gap magnitude, the superconducting gap at a site in the middle of the SC is used.

\begin{figure}[hbt]
    \centering
    \includegraphics{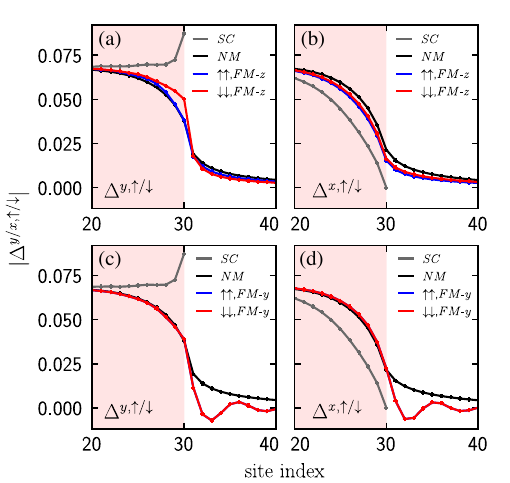}
    \caption{Spin-up and -down gap behaviour across the SC/FM interface shown for $p_y$ pairing in the left column (a) \& (c) and $p_x$ pairing in the right column (b) \& (d). Upper row (a) \& (b) shows behaviour for \textit{z}-polarized FM while lower row (c) \& (d) shows \textit{y}-polarized FM. The grey lines denotes the gap magnitude for a bare superconductor, the pink area denotes sites in the SC. The black lines show the gap magnitude in the case of an adjacent NM while red and blue denote spin-up and down gap behaviour in the presence of FM. Data obtained using $V_x=V_y=-0.85t$, $\mu=-0.5t$ with the same geometry as in Fig. \ref{fig: t_c vs thickness}. }
    \label{fig: d}
\end{figure}
\section{Results and discussion}
The normalized critical temperature $T_c$ for the SC/FM bilayer structure as a function of FM thickness depends on both the orbital and spin part of the superconducting order parameter as shown in Fig. (\ref{fig: t_c vs thickness}) for a system consisting of 30 real-space sites in the SC and 0-30 real-space sites in the FM. In the figures, $T_{c,0}$ is the critical temperature of the superconductor when it is not in proximity to any material.
The number of momentum modes was $N^y=121$ and an attractive interaction strength of $V_x=V_y=-0.85t$ was used. For the chosen parameters, the superconducting coherence length $\xi$ is estimated to be approximately 10 sites for $p_y$ pairing and around 15 sites in the case of $p_x$ pairing. $\xi$ is estimated by a visual inspection of the gap magnitude in the SC interior, see Fig. \ref{fig: coherence length}. The superconducting region of our system is thus 2-3 $\xi$ thick in the direction normal to the interface.  A chemical potential of $\mu=-0.5t$ was used to distinguish between spin-up and -down behaviour as the FM density-of-states (DOS) is spin-degenerate at the Fermi level independent of exchange field strength if one sets $\mu=0.0$.  The (a) subfigure shows the normalized $T_c$ in the case of $p_y$ pairing. i.e. pairing parallel to the SC/FM interface, while (b) presents the same results for $p_x$ pairing normal to the interface, both for the normal metal case as well as FM magnetized in \textit{y} and \textit{z} direction. For the spin-structure of the pairing considered in this paper, magnetization in \textit{z} and \textit{x} are equivalent. This can be understood in terms of the triplet $\boldsymbol{d}$-vector \cite{mackenzie_rmp_03} which for the superconducting pairing considered here has zero $x$ and $z$ components. We considered FM exchange field strengths ranging between $0.1-1.0t$. In the presentation of the results, given below, we will also compare $T_c$ in the triplet SC/FM case with $T_c$ in a triplet SC/NM case for a more complete understanding of the physics.

\subsection{$p_y$-wave pairing}

Considering first the dark grey lines in Fig. \ref{fig: t_c vs thickness}(a), the $T_c$ behaviour for $p_y$ pairing, we first observe that by adding normal metal (NM) layers to the SC (FM with $m=0.0$), the gap decays monotonously with layer number in an exponential decay-like manner. This is qualitatively understood as leakage of the superconducting pairing into the adjacent NM layers, causing a draining of the electrons that are part of the superconducting condensate which decreases the ``bulk" gap in the SC. As more layers are added, i.e. the NM becomes thicker, the effect of adding subsequent layers diminishes and $T_c$ eventually saturates. Replacing the NM by a FM magnetized along \textit{z}, i.e. along the spin quantization axis of the gap, this leakage is modified. The $T_c$ vs. FM thickness retains the exponential decay characteristic, but the asymptotic $T_c$ to which the system converges is shifted  and shows a clear dependence on the spin-orientation of the pairing. This is qualitatively understood by considering the DOS in the FM. At $\mu=-0.5t$, the DOS is spin-degenerate for $m=0.0$, but becomes spin-dependent under a non-zero exchange field. This causes an effective spin-dependent interface transmission as electrons with opposite spin have a differing density of available states to occupy in the FM. This causes the gap with spin orientation associated with the higher DOS to transmit more easily while the pairing of opposite spin electrons to a larger extent is confined to the SC interior. This is made clear in Fig. \ref{fig: d} where the $p_y$ gap magnitude is shown across the SC/FM interface for a z-polarized FM in the upper left panel. Compared to the NM case (grey) which is equal for both spin orientations, there is now a clear difference between spin-up and down pairing magnitude and the extent to which they leak into adjacent layers.  At $\mu=0.0$, the FM DOS remains spin-degenerate at $E=0.0$ also for finite exchange fields, albeit significantly lowered by the Zeeman splitting. The $T_c$ curves for spin-up and down gap coalesce for all magnetization strengths when we approach $\mu=0.0$. We finally point out that for the FM-z, $p_y$ pairing case, it is possible to have a combination of exchange field strength and chemical potential which causes the spin-dependent transmission of one spin species to surpass the NM case, causing a FM-induced lowering of $T_c$, at least for for the pairing type which is associated with a high DOS in the FM. We note however that for chemical potentials close to half-filling, the Zeeman splitting of the DOS usually results in a reduced or comparable DOS for both spin species. 

When the FM is polarized in \textit{y}, the behaviour of the $p_y$ pairing critical temperature changes significantly. Contrary to the exponential decay of the FM-z case, we now observe oscillatory behaviour of $T_c$ as function of FM thickness which is identical for the up- and down-polarized pairing (see Fig. \ref{fig: d}, lower left panel). If one considers the variation in exchange field strength (red curves in Fig. \ref{fig: t_c vs thickness} (a)), it is also evident that the oscillation frequency increases with field strength $m$. This behaviour is understood by recalling that as the spin quantization axis in the FM is rotated relative to the SC, the spin-up and -down gaps become non-diagonal in the new spin-basis and thus need to distribute across both spin-bands in the \textit{y}-polarized FM. This gives rise to pairing with a non-zero center-of-mass momentum in the FM, giving rise to the observed oscillations. The oscillatory behaviour of the gap is shown for $p_y$ pairing and a \textit{y}-polarized FM in Fig. \ref{fig: d} in the lower left panel. 

In other words, rotating the magnetization in-plane from $z$ to $y$ (\eg with an external field) changes the qualitative behavior of $T_c$ vs. the thickness of the FM from decaying to oscillatory. This is a unique signature of triplet pairing which would not occur either for BCS superconductors or even for high-$T_c$ $d$-wave superconductors since they are still spin-singlets. In those cases, only an oscillatory decaying behavior would be seen (albeit with a long oscillation length for a  weakly polarized FM). This is because spin-singlet pairs are rotationally invariant and thus do not respond differently to magnetizations in different directions. In the present case with intrinsic triplet superconductivity, this is no longer the case and rotating the magnetization causes $T_c$ to change.

\begin{figure}
    \centering
    \includegraphics{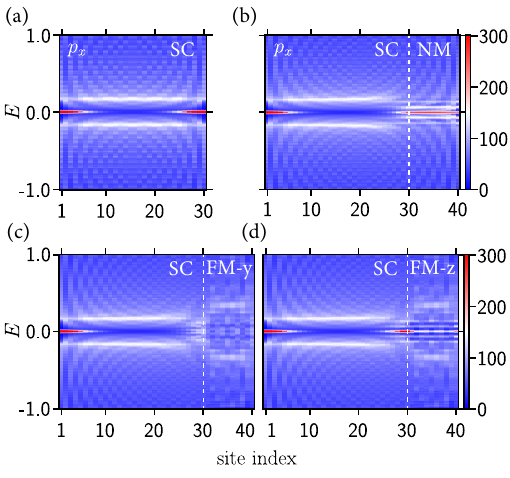}
    \caption{LDOS for the SC/NM and SC/FM heterostructures with $p_x$ pairing showing the isolated SC (a) and SC + 10 layers NM (b), FM-y (c) and FM-z (d), obtained using the same number of SC sites and momentum modes as in Fig. \ref{fig: t_c vs thickness}, but with $V_x=-1.00t$, $\mu=0.0$ to improve clarity. (a) From the isolated SC, it is evident that mid-gap Andreev bound states form at the vacuum boundaries (left and right side of SC). (b) When adding NM layers to the SC,  ABS at the SC/NM interface leaks into the adjacent layers, decreasing the presence of ABS in the SC interior. (c) In the presence of a \textit{y}-polarized FM the ABS is completely absent from the SC/FM interface while we observe that in the case of \textit{z}-polarized FM, the leakage of ABS is suppressed, compared to the NM case.}
    \label{fig: LDOS}
\end{figure}
\subsection{$p_x$-wave pairing: the role of Andreev bound states}
Keeping in mind the essential mechanisms discussed above, we now move on to the case of $p_x$ pairing, depicted in Fig. \ref{fig: t_c vs thickness} (b). Starting with the NM case, in contrast with the behaviour of the $p_y$ pairing, $T_c$ is now predicted to increase as the NM thickness increases. If the NM is replaced by a FM in \textit{z}, we note that the inverse exponential decay behaviour is preserved with a shifted asymptotic $T_c$. In order to understand the $T_c$ behaviour, we must consider a crucial difference between $p_y$ and $p_x$ pairing in our system, namely that the $p_x$ pairing, being of odd spatial symmetry and normal to the interface, is affected by the formation of Andreev bound states (ABS) at the vacuum boundary to the left in Fig. \ref{fig: system} as well as at the SC / FM interface \cite{Hu_PRL_94, TanakaKashiwaya_PRL_95}. 
ABS are mid-gap states, located within the superconducting gap. In regions where ABS are prominent, the superconducting gap is suppressed. In structures of limited thickness comparable to the superconducting coherence length, we can expect this interfacial gap suppression to have an effect on the gap throughout the superconducting region, causing a reduction in the critical temperature compared to the case of $p_y$ lacking ABS formation. With this in mind, we can now better grasp the $T_c$ behaviour observed in Fig. \ref{fig: t_c vs thickness} (b). As NM layers are added to the SC, the ABS originally confined to the SC interior of the SC / FM interface may leak into the adjacent layers, reducing their inhibitory effect on the gap in the SC. When the NM is replaced by a \textit{z}-polarized FM, this relocalization of the ABS from the SC to the adjacent layers is suppressed by the reduced DOS in the FM. This is shown in Fig. \ref{fig: d} (b) where we observe that the gap magnitude in the SC interior increases when NM layers are added (compare the grey and black lines), an effect which to some extent is suppressed when the NM is replaced by a FM-\textit{z} ($m=1.0t$) (red and blue lines in Fig. \ref{fig: d} (b)). In the case of a \textit{y}-polarized FM, we observe that the gap magnitude behaves like the gap in the presence of NM, but begins to oscillate in the FM, similarly as in the $p_y$ case. 

The effect of NM/FM layers on the ABS can be studied more closely by considering the LDOS, shown in Fig. \ref{fig: LDOS}. Here, LDOS is shown for an isolated SC as well as SC + 10 sites NM, FM-z and FM-y. From comparisons between Fig. \ref{fig: LDOS}(a) and (b), it follows that the presence of the NM layers suppresses ABS formation at the interface through a relocalization of the ABS into the NM layers as discussed above. Replacing the NM with a \textit{z}-polarized FM (Fig. \ref{fig: LDOS}(d)) reduces the suppressive effect on the ABS due to reduced interface transmission. The addition of \textit{y}-polarized FM layers has the opposite effect, the ABS states previously associated with the right SC boundary are completely absent in Fig. \ref{fig: LDOS}(b). This can be understood by recalling that we consider equal-spin pairing. ABS formation at the SC/FM interface due to the odd spatial symmetry of the order parameter necessarily has the same spin structure and is inhibited by the introduction of spin-flip scattering by the \textit{y}-polarized FM. Keeping this in mind, we can now understand why the presence of a \textit{y}-polarized FM gives a $T_c$ increase stronger than the NM case while a \textit{z}-polarized FM gives an increase smaller than the NM. 

Finally, we note that the effect of how the superconducting pairing symmetry affects the preferred stable magnetization direction was analyzed in \cite{gentile_prl_13}. In the present work, the magnetization direction is instead taken to be fixed, for instance by an internal magnetoanisotropy field which is greater than effective anisotropy induced by the superconducting state \cite{johnsen_prb_19}.

\section{Conclusion}
We have computed the critical temperature of a superconductor with intrinsic triplet pairing that is in proximity to a thin ferromagnetic layer. We show that the dependence of $T_c$ on the thickness $N_\text{layers}$ of the ferromagnet and its in-plane magnetization orientation are determined by both the spin and orbital part of the superconducting order parameter. Firstly, we find that $T_c$ either decays monotonically or oscillatory with $N_\text{layers}$ depending on the magnetization direction. This is fundamentally different from conventional BCS superconductors, where changing the in-plane magnetization direction leaves $T_c$ invariant. Secondly, we find that $T_c$ increases with $N_\text{layers}$ when the triplet superconducting order parameter features Andreev bound-states at the interface (such as $p_x$-wave pairing) whereas $T_c$ decreases with $N_\text{layers}$ when the order parameter does not (such as $p_y$-wave pairing). 
Since $T_c$ changes qualitatively depending on both the spin and orbital part of the superconducting order parameter, measurements of $T_c$ in a triplet SC/FM bilayer can be used to identify intrinsic spin-triplet pairing which is a highly sought after property in superconducting materials.

\acknowledgments
This work was supported by the Research Council of Norway through Grant No. 323766 and its Centres of Excellence funding scheme Grant No. 262633 “QuSpin”. Support from Sigma2 - the National Infrastructure for High-Performance Computing and Data Storage in Norway, project NN9577K, is acknowledged.

\bibliography{masterref.bib}

\begin{thebibliography}{23}%
\makeatletter
\providecommand \@ifxundefined [1]{%
 \@ifx{#1\undefined}
}%
\providecommand \@ifnum [1]{%
 \ifnum #1\expandafter \@firstoftwo
 \else \expandafter \@secondoftwo
 \fi
}%
\providecommand \@ifx [1]{%
 \ifx #1\expandafter \@firstoftwo
 \else \expandafter \@secondoftwo
 \fi
}%
\providecommand \natexlab [1]{#1}%
\providecommand \enquote  [1]{``#1''}%
\providecommand \bibnamefont  [1]{#1}%
\providecommand \bibfnamefont [1]{#1}%
\providecommand \citenamefont [1]{#1}%
\providecommand \href@noop [0]{\@secondoftwo}%
\providecommand \href [0]{\begingroup \@sanitize@url \@href}%
\providecommand \@href[1]{\@@startlink{#1}\@@href}%
\providecommand \@@href[1]{\endgroup#1\@@endlink}%
\providecommand \@sanitize@url [0]{\catcode `\\12\catcode `\$12\catcode `\&12\catcode `\#12\catcode `\^12\catcode `\_12\catcode `\%12\relax}%
\providecommand \@@startlink[1]{}%
\providecommand \@@endlink[0]{}%
\providecommand \url  [0]{\begingroup\@sanitize@url \@url }%
\providecommand \@url [1]{\endgroup\@href {#1}{\urlprefix }}%
\providecommand \urlprefix  [0]{URL }%
\providecommand \Eprint [0]{\href }%
\providecommand \doibase [0]{https://doi.org/}%
\providecommand \selectlanguage [0]{\@gobble}%
\providecommand \bibinfo  [0]{\@secondoftwo}%
\providecommand \bibfield  [0]{\@secondoftwo}%
\providecommand \translation [1]{[#1]}%
\providecommand \BibitemOpen [0]{}%
\providecommand \bibitemStop [0]{}%
\providecommand \bibitemNoStop [0]{.\EOS\space}%
\providecommand \EOS [0]{\spacefactor3000\relax}%
\providecommand \BibitemShut  [1]{\csname bibitem#1\endcsname}%
\let\auto@bib@innerbib\@empty
\bibitem [{\citenamefont {Buzdin}(2005)}]{buzdin_rmp_05}%
  \BibitemOpen
  \bibfield  {author} {\bibinfo {author} {\bibfnamefont {A.~I.}\ \bibnamefont {Buzdin}},\ }\href {https://journals.aps.org/rmp/abstract/10.1103/RevModPhys.77.935} {\bibfield  {journal} {\bibinfo  {journal} {Rev. Mod. Phys.}\ }\textbf {\bibinfo {volume} {77}},\ \bibinfo {pages} {935} (\bibinfo {year} {2005})}\BibitemShut {NoStop}%
\bibitem [{\citenamefont {Bergeret}\ \emph {et~al.}(2005)\citenamefont {Bergeret}, \citenamefont {Volkov},\ and\ \citenamefont {Efetov}}]{bergeret_rmp_05}%
  \BibitemOpen
  \bibfield  {author} {\bibinfo {author} {\bibfnamefont {F.~S.}\ \bibnamefont {Bergeret}}, \bibinfo {author} {\bibfnamefont {A.~F.}\ \bibnamefont {Volkov}},\ and\ \bibinfo {author} {\bibfnamefont {K.~B.}\ \bibnamefont {Efetov}},\ }\href {https://doi.org/10.1103/RevModPhys.77.1321} {\bibfield  {journal} {\bibinfo  {journal} {Rev. Mod. Phys.}\ }\textbf {\bibinfo {volume} {77}},\ \bibinfo {pages} {1321} (\bibinfo {year} {2005})}\BibitemShut {NoStop}%
\bibitem [{\citenamefont {Bergeret}\ \emph {et~al.}(2001)\citenamefont {Bergeret}, \citenamefont {Volkov},\ and\ \citenamefont {Efetov}}]{bergeret_prl_01}%
  \BibitemOpen
  \bibfield  {author} {\bibinfo {author} {\bibfnamefont {F.~S.}\ \bibnamefont {Bergeret}}, \bibinfo {author} {\bibfnamefont {A.~F.}\ \bibnamefont {Volkov}},\ and\ \bibinfo {author} {\bibfnamefont {K.~B.}\ \bibnamefont {Efetov}},\ }\href {https://doi.org/10.1103/PhysRevLett.86.4096} {\bibfield  {journal} {\bibinfo  {journal} {Phys. Rev. Lett.}\ }\textbf {\bibinfo {volume} {86}},\ \bibinfo {pages} {4096} (\bibinfo {year} {2001})}\BibitemShut {NoStop}%
\bibitem [{\citenamefont {Keizer}\ \emph {et~al.}(2006)\citenamefont {Keizer}, \citenamefont {Goennenwein}, \citenamefont {Klapwijk}, \citenamefont {Miao},\ and\ \citenamefont {Gupta}}]{keizer_nature_06}%
  \BibitemOpen
  \bibfield  {author} {\bibinfo {author} {\bibfnamefont {R.~S.}\ \bibnamefont {Keizer}}, \bibinfo {author} {\bibfnamefont {S.~T.~B.}\ \bibnamefont {Goennenwein}}, \bibinfo {author} {\bibfnamefont {T.~M.}\ \bibnamefont {Klapwijk}}, \bibinfo {author} {\bibfnamefont {G.}~\bibnamefont {Miao}},\ and\ \bibinfo {author} {\bibfnamefont {A.}~\bibnamefont {Gupta}},\ }\href {https://www.nature.com/articles/nature04499} {\bibfield  {journal} {\bibinfo  {journal} {Nature}\ }\textbf {\bibinfo {volume} {439}},\ \bibinfo {pages} {825} (\bibinfo {year} {2006})}\BibitemShut {NoStop}%
\bibitem [{\citenamefont {Khaire}\ \emph {et~al.}(2010)\citenamefont {Khaire}, \citenamefont {Khasawneh}, \citenamefont {Pratt},\ and\ \citenamefont {Birge}}]{khaire_prl_10}%
  \BibitemOpen
  \bibfield  {author} {\bibinfo {author} {\bibfnamefont {T.~S.}\ \bibnamefont {Khaire}}, \bibinfo {author} {\bibfnamefont {M.~A.}\ \bibnamefont {Khasawneh}}, \bibinfo {author} {\bibfnamefont {W.~P.}\ \bibnamefont {Pratt}},\ and\ \bibinfo {author} {\bibfnamefont {N.~O.}\ \bibnamefont {Birge}},\ }\href {https://journals.aps.org/prl/abstract/10.1103/PhysRevLett.104.137002} {\bibfield  {journal} {\bibinfo  {journal} {Phys. Rev. Lett.}\ }\textbf {\bibinfo {volume} {104}},\ \bibinfo {pages} {137002} (\bibinfo {year} {2010})}\BibitemShut {NoStop}%
\bibitem [{\citenamefont {Robinson}\ \emph {et~al.}(2010)\citenamefont {Robinson}, \citenamefont {Witt},\ and\ \citenamefont {Blamire}}]{robinson_science_10}%
  \BibitemOpen
  \bibfield  {author} {\bibinfo {author} {\bibfnamefont {J.~W.~A.}\ \bibnamefont {Robinson}}, \bibinfo {author} {\bibfnamefont {J.~D.~S.}\ \bibnamefont {Witt}},\ and\ \bibinfo {author} {\bibfnamefont {M.~G.}\ \bibnamefont {Blamire}},\ }\href {https://doi.org/10.1126/science.1189246} {\bibfield  {journal} {\bibinfo  {journal} {Science}\ }\textbf {\bibinfo {volume} {329}},\ \bibinfo {pages} {59} (\bibinfo {year} {2010})}\BibitemShut {NoStop}%
\bibitem [{\citenamefont {Eschrig}(2015)}]{Eschrig2015a}%
  \BibitemOpen
  \bibfield  {author} {\bibinfo {author} {\bibfnamefont {M.}~\bibnamefont {Eschrig}},\ }\href {https://doi.org/10.1088/0034-4885/78/10/104501} {\bibfield  {journal} {\bibinfo  {journal} {Rep. Prog. Phys.}\ }\textbf {\bibinfo {volume} {78}},\ \bibinfo {pages} {104501} (\bibinfo {year} {2015})}\BibitemShut {NoStop}%
\bibitem [{\citenamefont {Linder}\ and\ \citenamefont {Robinson}(2015)}]{linder_nphys_15}%
  \BibitemOpen
  \bibfield  {author} {\bibinfo {author} {\bibfnamefont {J.}~\bibnamefont {Linder}}\ and\ \bibinfo {author} {\bibfnamefont {J.~W.~A.}\ \bibnamefont {Robinson}},\ }\href {https://doi.org/10.1038/nphys3242} {\bibfield  {journal} {\bibinfo  {journal} {Nat. Phys.}\ }\textbf {\bibinfo {volume} {11}},\ \bibinfo {pages} {307} (\bibinfo {year} {2015})}\BibitemShut {NoStop}%
\bibitem [{\citenamefont {Saxena}\ \emph {et~al.}(2000)\citenamefont {Saxena}, \citenamefont {Agarwal}, \citenamefont {Ahilan}, \citenamefont {Grosche}, \citenamefont {Haselwimmer}, \citenamefont {Steiner}, \citenamefont {Pugh}, \citenamefont {Walker}, \citenamefont {Julian}, \citenamefont {Monthoux}, \citenamefont {Lonzarich}, \citenamefont {Huxley}, \citenamefont {Sheikin}, \citenamefont {Braithwaite},\ and\ \citenamefont {Flouquet}}]{Saxena2000}%
  \BibitemOpen
  \bibfield  {author} {\bibinfo {author} {\bibfnamefont {S.~S.}\ \bibnamefont {Saxena}}, \bibinfo {author} {\bibfnamefont {P.}~\bibnamefont {Agarwal}}, \bibinfo {author} {\bibfnamefont {K.}~\bibnamefont {Ahilan}}, \bibinfo {author} {\bibfnamefont {F.~M.}\ \bibnamefont {Grosche}}, \bibinfo {author} {\bibfnamefont {R.~K.~W.}\ \bibnamefont {Haselwimmer}}, \bibinfo {author} {\bibfnamefont {M.~J.}\ \bibnamefont {Steiner}}, \bibinfo {author} {\bibfnamefont {E.}~\bibnamefont {Pugh}}, \bibinfo {author} {\bibfnamefont {I.~R.}\ \bibnamefont {Walker}}, \bibinfo {author} {\bibfnamefont {S.~R.}\ \bibnamefont {Julian}}, \bibinfo {author} {\bibfnamefont {P.}~\bibnamefont {Monthoux}}, \bibinfo {author} {\bibfnamefont {G.~G.}\ \bibnamefont {Lonzarich}}, \bibinfo {author} {\bibfnamefont {A.}~\bibnamefont {Huxley}}, \bibinfo {author} {\bibfnamefont {I.}~\bibnamefont {Sheikin}}, \bibinfo {author} {\bibfnamefont {D.}~\bibnamefont {Braithwaite}},\ and\ \bibinfo {author} {\bibfnamefont {J.}~\bibnamefont {Flouquet}},\ }\href
  {https://doi.org/10.1038/35020500} {\bibfield  {journal} {\bibinfo  {journal} {Nature}\ }\textbf {\bibinfo {volume} {406}},\ \bibinfo {pages} {587–592} (\bibinfo {year} {2000})}\BibitemShut {NoStop}%
\bibitem [{\citenamefont {Aoki}\ \emph {et~al.}(2001)\citenamefont {Aoki}, \citenamefont {Huxley}, \citenamefont {Ressouche}, \citenamefont {Braithwaite}, \citenamefont {Flouquet}, \citenamefont {Brison}, \citenamefont {Lhotel},\ and\ \citenamefont {Paulsen}}]{Aoki2001}%
  \BibitemOpen
  \bibfield  {author} {\bibinfo {author} {\bibfnamefont {D.}~\bibnamefont {Aoki}}, \bibinfo {author} {\bibfnamefont {A.}~\bibnamefont {Huxley}}, \bibinfo {author} {\bibfnamefont {E.}~\bibnamefont {Ressouche}}, \bibinfo {author} {\bibfnamefont {D.}~\bibnamefont {Braithwaite}}, \bibinfo {author} {\bibfnamefont {J.}~\bibnamefont {Flouquet}}, \bibinfo {author} {\bibfnamefont {J.-P.}\ \bibnamefont {Brison}}, \bibinfo {author} {\bibfnamefont {E.}~\bibnamefont {Lhotel}},\ and\ \bibinfo {author} {\bibfnamefont {C.}~\bibnamefont {Paulsen}},\ }\href {https://doi.org/10.1038/35098048} {\bibfield  {journal} {\bibinfo  {journal} {Nature}\ }\textbf {\bibinfo {volume} {413}},\ \bibinfo {pages} {613–616} (\bibinfo {year} {2001})}\BibitemShut {NoStop}%
\bibitem [{\citenamefont {Aoki}\ \emph {et~al.}(2011)\citenamefont {Aoki}, \citenamefont {Hardy}, \citenamefont {Miyake}, \citenamefont {Taufour}, \citenamefont {Matsuda},\ and\ \citenamefont {Flouquet}}]{Aoki2011}%
  \BibitemOpen
  \bibfield  {author} {\bibinfo {author} {\bibfnamefont {D.}~\bibnamefont {Aoki}}, \bibinfo {author} {\bibfnamefont {F.}~\bibnamefont {Hardy}}, \bibinfo {author} {\bibfnamefont {A.}~\bibnamefont {Miyake}}, \bibinfo {author} {\bibfnamefont {V.}~\bibnamefont {Taufour}}, \bibinfo {author} {\bibfnamefont {T.~D.}\ \bibnamefont {Matsuda}},\ and\ \bibinfo {author} {\bibfnamefont {J.}~\bibnamefont {Flouquet}},\ }\href {https://doi.org/10.1016/j.crhy.2011.04.007} {\bibfield  {journal} {\bibinfo  {journal} {C. R. Physique}\ }\textbf {\bibinfo {volume} {12}},\ \bibinfo {pages} {573} (\bibinfo {year} {2011})}\BibitemShut {NoStop}%
\bibitem [{\citenamefont {Huy}\ \emph {et~al.}(2007)\citenamefont {Huy}, \citenamefont {Gasparini}, \citenamefont {de~Nijs}, \citenamefont {Huang}, \citenamefont {Klaasse}, \citenamefont {Gortenmulder}, \citenamefont {de~Visser}, \citenamefont {Hamann}, \citenamefont {G\"orlach},\ and\ \citenamefont {L\"ohneysen}}]{Huy_prl2007}%
  \BibitemOpen
  \bibfield  {author} {\bibinfo {author} {\bibfnamefont {N.~T.}\ \bibnamefont {Huy}}, \bibinfo {author} {\bibfnamefont {A.}~\bibnamefont {Gasparini}}, \bibinfo {author} {\bibfnamefont {D.~E.}\ \bibnamefont {de~Nijs}}, \bibinfo {author} {\bibfnamefont {Y.}~\bibnamefont {Huang}}, \bibinfo {author} {\bibfnamefont {J.~C.~P.}\ \bibnamefont {Klaasse}}, \bibinfo {author} {\bibfnamefont {T.}~\bibnamefont {Gortenmulder}}, \bibinfo {author} {\bibfnamefont {A.}~\bibnamefont {de~Visser}}, \bibinfo {author} {\bibfnamefont {A.}~\bibnamefont {Hamann}}, \bibinfo {author} {\bibfnamefont {T.}~\bibnamefont {G\"orlach}},\ and\ \bibinfo {author} {\bibfnamefont {H.~v.}\ \bibnamefont {L\"ohneysen}},\ }\href {https://doi.org/10.1103/PhysRevLett.99.067006} {\bibfield  {journal} {\bibinfo  {journal} {Phys. Rev. Lett.}\ }\textbf {\bibinfo {volume} {99}},\ \bibinfo {pages} {067006} (\bibinfo {year} {2007})}\BibitemShut {NoStop}%
\bibitem [{\citenamefont {Cirillo}\ \emph {et~al.}(2005)\citenamefont {Cirillo}, \citenamefont {Prischepa}, \citenamefont {Salvato}, \citenamefont {Attanasio}, \citenamefont {Hesselberth},\ and\ \citenamefont {Aarts}}]{cirillo_prb_05}%
  \BibitemOpen
  \bibfield  {author} {\bibinfo {author} {\bibfnamefont {C.}~\bibnamefont {Cirillo}}, \bibinfo {author} {\bibfnamefont {S.~L.}\ \bibnamefont {Prischepa}}, \bibinfo {author} {\bibfnamefont {M.}~\bibnamefont {Salvato}}, \bibinfo {author} {\bibfnamefont {C.}~\bibnamefont {Attanasio}}, \bibinfo {author} {\bibfnamefont {M.}~\bibnamefont {Hesselberth}},\ and\ \bibinfo {author} {\bibfnamefont {J.}~\bibnamefont {Aarts}},\ }\href {https://doi.org/10.1103/PhysRevB.72.144511} {\bibfield  {journal} {\bibinfo  {journal} {Phys. Rev. B}\ }\textbf {\bibinfo {volume} {72}},\ \bibinfo {pages} {144511} (\bibinfo {year} {2005})}\BibitemShut {NoStop}%
\bibitem [{\citenamefont {Zdravkov}\ \emph {et~al.}(2006)\citenamefont {Zdravkov}, \citenamefont {Sidorenko}, \citenamefont {Obermeier}, \citenamefont {Gsell}, \citenamefont {Schreck}, \citenamefont {M\"uller}, \citenamefont {Horn}, \citenamefont {Tidecks},\ and\ \citenamefont {Tagirov}}]{zdravkov_prl_06}%
  \BibitemOpen
  \bibfield  {author} {\bibinfo {author} {\bibfnamefont {V.}~\bibnamefont {Zdravkov}}, \bibinfo {author} {\bibfnamefont {A.}~\bibnamefont {Sidorenko}}, \bibinfo {author} {\bibfnamefont {G.}~\bibnamefont {Obermeier}}, \bibinfo {author} {\bibfnamefont {S.}~\bibnamefont {Gsell}}, \bibinfo {author} {\bibfnamefont {M.}~\bibnamefont {Schreck}}, \bibinfo {author} {\bibfnamefont {C.}~\bibnamefont {M\"uller}}, \bibinfo {author} {\bibfnamefont {S.}~\bibnamefont {Horn}}, \bibinfo {author} {\bibfnamefont {R.}~\bibnamefont {Tidecks}},\ and\ \bibinfo {author} {\bibfnamefont {L.~R.}\ \bibnamefont {Tagirov}},\ }\href {https://doi.org/10.1103/PhysRevLett.97.057004} {\bibfield  {journal} {\bibinfo  {journal} {Phys. Rev. Lett.}\ }\textbf {\bibinfo {volume} {97}},\ \bibinfo {pages} {057004} (\bibinfo {year} {2006})}\BibitemShut {NoStop}%
\bibitem [{\citenamefont {Leksin}\ \emph {et~al.}(2012)\citenamefont {Leksin}, \citenamefont {Garif'yanov}, \citenamefont {Garifullin}, \citenamefont {Fominov}, \citenamefont {Schumann}, \citenamefont {Krupskaya}, \citenamefont {Kataev}, \citenamefont {Schmidt},\ and\ \citenamefont {B\"uchner}}]{leksin_prl_12}%
  \BibitemOpen
  \bibfield  {author} {\bibinfo {author} {\bibfnamefont {P.~V.}\ \bibnamefont {Leksin}}, \bibinfo {author} {\bibfnamefont {N.~N.}\ \bibnamefont {Garif'yanov}}, \bibinfo {author} {\bibfnamefont {I.~A.}\ \bibnamefont {Garifullin}}, \bibinfo {author} {\bibfnamefont {Y.~V.}\ \bibnamefont {Fominov}}, \bibinfo {author} {\bibfnamefont {J.}~\bibnamefont {Schumann}}, \bibinfo {author} {\bibfnamefont {Y.}~\bibnamefont {Krupskaya}}, \bibinfo {author} {\bibfnamefont {V.}~\bibnamefont {Kataev}}, \bibinfo {author} {\bibfnamefont {O.~G.}\ \bibnamefont {Schmidt}},\ and\ \bibinfo {author} {\bibfnamefont {B.}~\bibnamefont {B\"uchner}},\ }\href@noop {} {\bibfield  {journal} {\bibinfo  {journal} {Phys. Rev. Lett.}\ }\textbf {\bibinfo {volume} {109}},\ \bibinfo {pages} {057005} (\bibinfo {year} {2012})}\BibitemShut {NoStop}%
\bibitem [{\citenamefont {Gu}\ \emph {et~al.}(2014)\citenamefont {Gu}, \citenamefont {Robinson}, \citenamefont {Bianchetti}, \citenamefont {Stelmashenko}, \citenamefont {Astill}, \citenamefont {Grosche}, \citenamefont {MacManus-Driscoll},\ and\ \citenamefont {Blamire}}]{gu_apl_14}%
  \BibitemOpen
  \bibfield  {author} {\bibinfo {author} {\bibfnamefont {Y.}~\bibnamefont {Gu}}, \bibinfo {author} {\bibfnamefont {J.~W.~A.}\ \bibnamefont {Robinson}}, \bibinfo {author} {\bibfnamefont {M.}~\bibnamefont {Bianchetti}}, \bibinfo {author} {\bibfnamefont {N.~A.}\ \bibnamefont {Stelmashenko}}, \bibinfo {author} {\bibfnamefont {D.}~\bibnamefont {Astill}}, \bibinfo {author} {\bibfnamefont {F.~M.}\ \bibnamefont {Grosche}}, \bibinfo {author} {\bibfnamefont {J.~L.}\ \bibnamefont {MacManus-Driscoll}},\ and\ \bibinfo {author} {\bibfnamefont {M.~G.}\ \bibnamefont {Blamire}},\ }\href@noop {} {\bibfield  {journal} {\bibinfo  {journal} {APL Materials}\ }\textbf {\bibinfo {volume} {2}},\ \bibinfo {pages} {046103} (\bibinfo {year} {2014})}\BibitemShut {NoStop}%
\bibitem [{\citenamefont {Ouassou}(2015)}]{ouassou2015density}%
  \BibitemOpen
  \bibfield  {author} {\bibinfo {author} {\bibfnamefont {J.~A.}\ \bibnamefont {Ouassou}},\ }\emph {\bibinfo {title} {Density of states and critical temperature in superconductor/ferromagnet structures with spin-orbit coupling}},\ \href@noop {} {Master's thesis},\ \bibinfo  {school} {NTNU} (\bibinfo {year} {2015})\BibitemShut {NoStop}%
\bibitem [{\citenamefont {Johnsen}\ \emph {et~al.}(2020)\citenamefont {Johnsen}, \citenamefont {Svalland},\ and\ \citenamefont {Linder}}]{Johnsen2020:PRL}%
  \BibitemOpen
  \bibfield  {author} {\bibinfo {author} {\bibfnamefont {L.~G.}\ \bibnamefont {Johnsen}}, \bibinfo {author} {\bibfnamefont {K.}~\bibnamefont {Svalland}},\ and\ \bibinfo {author} {\bibfnamefont {J.}~\bibnamefont {Linder}},\ }\href {https://doi.org/10.1103/PhysRevLett.125.107002} {\bibfield  {journal} {\bibinfo  {journal} {Phys. Rev. Lett.}\ }\textbf {\bibinfo {volume} {125}},\ \bibinfo {pages} {107002} (\bibinfo {year} {2020})}\BibitemShut {NoStop}%
\bibitem [{\citenamefont {Mackenzie}\ and\ \citenamefont {Maeno}(2003)}]{mackenzie_rmp_03}%
  \BibitemOpen
  \bibfield  {author} {\bibinfo {author} {\bibfnamefont {A.~P.}\ \bibnamefont {Mackenzie}}\ and\ \bibinfo {author} {\bibfnamefont {Y.}~\bibnamefont {Maeno}},\ }\href {https://doi.org/10.1103/RevModPhys.75.657} {\bibfield  {journal} {\bibinfo  {journal} {Rev. Mod. Phys.}\ }\textbf {\bibinfo {volume} {75}},\ \bibinfo {pages} {657} (\bibinfo {year} {2003})}\BibitemShut {NoStop}%
\bibitem [{\citenamefont {Hu}(1994)}]{Hu_PRL_94}%
  \BibitemOpen
  \bibfield  {author} {\bibinfo {author} {\bibfnamefont {C.-R.}\ \bibnamefont {Hu}},\ }\href {https://doi.org/10.1103/PhysRevLett.72.1526} {\bibfield  {journal} {\bibinfo  {journal} {Phys. Rev. Lett.}\ }\textbf {\bibinfo {volume} {72}},\ \bibinfo {pages} {1526} (\bibinfo {year} {1994})}\BibitemShut {NoStop}%
\bibitem [{\citenamefont {Tanaka}\ and\ \citenamefont {Kashiwaya}(1995)}]{TanakaKashiwaya_PRL_95}%
  \BibitemOpen
  \bibfield  {author} {\bibinfo {author} {\bibfnamefont {Y.}~\bibnamefont {Tanaka}}\ and\ \bibinfo {author} {\bibfnamefont {S.}~\bibnamefont {Kashiwaya}},\ }\href {https://doi.org/10.1103/PhysRevLett.74.3451} {\bibfield  {journal} {\bibinfo  {journal} {Phys. Rev. Lett.}\ }\textbf {\bibinfo {volume} {74}},\ \bibinfo {pages} {3451} (\bibinfo {year} {1995})}\BibitemShut {NoStop}%
\bibitem [{\citenamefont {Gentile}\ \emph {et~al.}(2013)\citenamefont {Gentile}, \citenamefont {Cuoco}, \citenamefont {Romano}, \citenamefont {Noce}, \citenamefont {Manske},\ and\ \citenamefont {Brydon}}]{gentile_prl_13}%
  \BibitemOpen
  \bibfield  {author} {\bibinfo {author} {\bibfnamefont {P.}~\bibnamefont {Gentile}}, \bibinfo {author} {\bibfnamefont {M.}~\bibnamefont {Cuoco}}, \bibinfo {author} {\bibfnamefont {A.}~\bibnamefont {Romano}}, \bibinfo {author} {\bibfnamefont {C.}~\bibnamefont {Noce}}, \bibinfo {author} {\bibfnamefont {D.}~\bibnamefont {Manske}},\ and\ \bibinfo {author} {\bibfnamefont {P.~M.~R.}\ \bibnamefont {Brydon}},\ }\href {https://doi.org/10.1103/PhysRevLett.111.097003} {\bibfield  {journal} {\bibinfo  {journal} {Phys. Rev. Lett.}\ }\textbf {\bibinfo {volume} {111}},\ \bibinfo {pages} {097003} (\bibinfo {year} {2013})}\BibitemShut {NoStop}%
\bibitem [{\citenamefont {Johnsen}\ \emph {et~al.}(2019)\citenamefont {Johnsen}, \citenamefont {Banerjee},\ and\ \citenamefont {Linder}}]{johnsen_prb_19}%
  \BibitemOpen
  \bibfield  {author} {\bibinfo {author} {\bibfnamefont {L.~G.}\ \bibnamefont {Johnsen}}, \bibinfo {author} {\bibfnamefont {N.}~\bibnamefont {Banerjee}},\ and\ \bibinfo {author} {\bibfnamefont {J.}~\bibnamefont {Linder}},\ }\href@noop {} {\bibfield  {journal} {\bibinfo  {journal} {Phys. Rev. B}\ }\textbf {\bibinfo {volume} {99}},\ \bibinfo {pages} {134516} (\bibinfo {year} {2019})}\BibitemShut {NoStop}%
\end{thebibliography}%

\begin{widetext}

\clearpage
\appendix

\section{$p$-wave term in mean field theory} \label{App: p-wave}
We consider now the nearest-neighbour attractive interaction modelled by the term
\begin{equation}
    \sum_{\langle i, j \rangle, \sigma} V_{ij} n_{i,\sigma}n_{j,\sigma}
\end{equation}
in the Hamiltonian. We assume now that the site coordinate caries two components, $i=(i_x, i_y)$ where \textit{x} is the direction normal to the interface and \textit{y} the parallel direction. We may then introduce Fourier-transformed operators in the latter direction in order to take full advantage of the periodicity of the system. We proceed by rewriting the Hamiltonian using these transformed operators, assuming the attractive interaction to be isotropic, $V_{ij}=V\leq 0$. 
\begin{align}
    \sum_{\langle i, j \rangle, \sigma} V_{ij} n_{i,\sigma}n_{j,\sigma}&=V\sum_{(i_x, i_y), \boldsymbol{\delta}, \sigma}  n_{(i_x, i_y),\sigma}n_{(i_x+\delta_x, i_y+\delta_y),\sigma} \\
    &=V\sum_{(i_x, i_y), \boldsymbol{\delta}, \sigma} \cc[]{(i_x, i_y),\sigma}\ca[]{(i_x, i_y), \sigma}\cc[]{(i_x+\delta_x, i_y+\delta_y), \sigma}\ca[]{(i_x+\delta_x, i_y+\delta_y), \sigma}
\end{align}
where $\delta_{x/y}$ are the \textit{x} and \textit{y} components of the nearest neighbour vector. We now introduce Fourier-transformed operators in the \textit{y} direction: 
\begin{align}
    V\sum_{(i_x, i_y), \boldsymbol{\delta}, \sigma}& \cc[]{(i_x, i_y),\sigma}\ca[]{(i_x, i_y), \sigma}\cc[]{(i_x+\delta_x, i_y+\delta_y), \sigma}\ca[]{(i_x+\delta_x, i_y+\delta_y), \sigma}\\ &=\frac{V}{(N^y)^2}\sum_{(i_x, i_y), \boldsymbol{\delta}, \sigma}\sum_{k_y^1, k_y^2, k_y^3, k_y^4} \cc[]{i_x, k_y^1,\sigma}\ca[]{i_x, k_y^2, \sigma}\cc[]{i_x+\delta_x, k_y^3, \sigma}\ca[]{i_x+\delta_x, k_y^4, \sigma}e^{-i(k_y^1 - k_y^2 + k_y^3 - k_y^4)r_{i_y}}e^{-i(k_y^3-k_y^4)\delta_y} \\
    &=\frac{V}{N^y} \sum_{i_x, \boldsymbol{\delta},\sigma}\sum_{k_y^2, k_y^3, k_y^4}\cc[]{i_x, k_y^2 - k_y^3 + k_y^4,\sigma}\ca[]{i_x, k_y^2, \sigma}\cc[]{i_x+\delta_x, k_y^3, \sigma}\ca[]{i_x+\delta_x, k_y^4, \sigma}e^{-i(k_y^3-k_y^4)\delta_y} 
\end{align}
Renaming the momentum variables now as $k_y^2=k$, $k_y^4= k'$ and $k_y^3=k'-q$, ignoring the \textit{y} subscripts from now on,  we may rewrite the above as 
\begin{align}
    \frac{V}{N^y} &\sum_{i_x, \boldsymbol{\delta},\sigma}\sum_{k, k', q}\cc[]{i_x, k + q,\sigma}\ca[]{i_x, k, \sigma}\cc[]{i_x+\delta_x, k'-q, \sigma}\ca[]{i_x+\delta_x, k', \sigma}e^{iq\delta_y}\\\qquad&=\frac{V}{N^y} \sum_{i_x, \boldsymbol{\delta},\sigma}\sum_{k, k', q}\cc[]{i_x+\delta_x, k'-q, \sigma}\cc[]{i_x, k + q,\sigma}\ca[]{i_x, k, \sigma}\ca[]{i_x+\delta_x, k', \sigma}e^{iq\delta_y}
\end{align}
We now proceed by using that the attractive interaction only facilitates scattering between states located in a thin shell around the Fermi surface. By considering the kinematics, it follows that the scattering phase space is maximized by considering the case where $k'=-k$. In the following, we shall keep only such terms. We can then rewrite the expression as 
\begin{align}
\frac{V}{N^y} &\sum_{i_x, \boldsymbol{\delta},\sigma}\sum_{k, q}\cc[]{i_x+\delta_x, -k-q, \sigma}\cc[]{i_x, k + q,\sigma}\ca[]{i_x, k, \sigma}\ca[]{i_x+\delta_x, -k, \sigma}e^{iq\delta_y} \\ \qquad &=\frac{V}{N^y} \sum_{i_x, \boldsymbol{\delta},\sigma}\sum_{k, k'}\cc[]{i_x+\delta_x, -k', \sigma}\cc[]{i_x,k',\sigma}\ca[]{i_x, k, \sigma}\ca[]{i_x+\delta_x, -k, \sigma}e^{i(k'-k)\delta_y}
\end{align}
where we have introduced a final variable $k'=k+q$. We now introduce the mean fields $\langle \cc[]{i_x+\delta_x, -k', \sigma}\cc[]{i_x,k',\sigma}\rangle$, $\langle \ca[]{i_x, k, \sigma}\ca[]{i_x+\delta_x, -k, \sigma}\rangle$ and rewrite the above to first order in the deviation from the mean fields, 
\begin{align}
    &\frac{V}{N^y}=\sum_{i_x, \boldsymbol{\delta},\sigma}\sum_{k, k'}\cc[]{i_x+\delta_x, -k', \sigma}\cc[]{i_x,k',\sigma}\langle \ca[]{i_x, k, \sigma}\ca[]{i_x+\delta_x, -k, \sigma}\rangle e^{i(k'-k)\delta_y} + \text{h.c.} \\
    \qquad&=\sum_{i_x,\sigma}\sum_{k}\cc[]{i_x+\hat{x}, -k, \sigma}\cc[]{i_x,k,\sigma}\Delta_{i}^{\sigma,\hat{x}}+\cc[]{i_x-\hat{x}, -k, \sigma}\cc[]{i_x,k,\sigma}\Delta_{i}^{\sigma,-\hat{x}} + \cc[]{i_x, -k, \sigma}\cc[]{i_x,k,\sigma}\Delta_{k}^{\sigma,\hat{\boldsymbol{y}}} + \text{h.c.}
\end{align}
where we have defined the gaps
\begin{align}
    \Delta_{i}^{\sigma,\pm\hat{x}}&=\frac{V_x}{N^{y}}\sum_{k'} F_{i,k'}^{\sigma, \pm\hat{x}} \\
    \Delta_{i, k}^{\sigma, \hat{y}}&= \frac{V_y}{N^{y}}\sum_{k'}2\cos(k'-k) F_{i,k'}^{\sigma, \hat{y}}, 
\end{align} 
and the pairing amplitudes 
\begin{align}
F_{i,k'}^{\sigma,\pm\hat{x}}&=\big\langle \ca[]{i, k', \sigma}\ca[]{i\pm\hat{x}, -k', \sigma}\big\rangle \\[10pt] F_{i,k'}^{\sigma,\hat{y}}&=\langle \ca[]{i, k', \sigma}\ca[]{i, -k', \sigma}\big\rangle,
\end{align}
We can write now write the contribution to the Hamiltonian from the $p$-wave coupling as 
\begin{equation}
    H=E_0 + \frac{1}{2}\sum_{i_x, j_x}B_{i_x}^\dagger h_{i_x, j_x} B_{j_x} 
\end{equation}
where we have for $i_x = j_x$, 
\begin{equation}
    h_{i_x, i_x} = \sum_{k}\begin{pmatrix}
        \cc[]{i_x, k, \uparrow} \\ \cc[]{i_x, k, \downarrow} \\
        \ca[]{i_x, -k, \uparrow} \\ \ca[]{i_x, -k, \downarrow}
    \end{pmatrix}^T \begin{bmatrix}
        0 & 0 & -2\Delta_{i, k}^{\uparrow, \hat{y}} & 0 \\
        0 & 0 & 0 & -2\Delta_{i, k}^{\downarrow, \hat{y}} \\ 
        -2(\Delta_{i, k}^{\uparrow, \hat{y}})^\dagger & 0 & 0 & 0 \\
        0 & -2(\Delta_{i, k}^{\downarrow, \hat{y}})^\dagger & 0 & 0
    \end{bmatrix}\begin{pmatrix}
        \ca[]{i_x, k, \uparrow} \\ \ca[]{i_x, k, \downarrow} \\
        \cc[]{i_x, -k, \uparrow} \\ \cc[]{i_x, -k, \downarrow}
    \end{pmatrix}
\end{equation}
and $j_x=i_x\pm \hat{x}$,
\begin{equation}
    h_{i_x, i_x\pm\hat{x}} = \sum_{k}\begin{pmatrix}
        \cc[]{i_x, k, \uparrow} \\ \cc[]{i_x, k, \downarrow} \\
        \ca[]{i_x, -k, \uparrow} \\ \ca[]{i_x, -k, \downarrow}
    \end{pmatrix}^T \begin{bmatrix}
        0 & 0 & -2\Delta_{i}^{\uparrow, \pm\hat{x}} & 0 \\
        0 & 0 & 0 & -2\Delta_{i}^{\downarrow, \pm\hat{x}} \\ 
        2(\Delta_{i}^{\uparrow, \pm\hat{x}})^\dagger & 0 & 0 & 0 \\
        0 & 2(\Delta_{i}^{\downarrow, \pm\hat{x}})^\dagger & 0 & 0
    \end{bmatrix}\begin{pmatrix}
        \ca[]{i_x\pm\hat{x}, k, \uparrow} \\ \ca[]{i_x\pm\hat{x}, k, \downarrow} \\
        \cc[]{i_x\pm\hat{x}, -k, \uparrow} \\ \cc[]{i_x\pm\hat{x}, -k, \downarrow}
    \end{pmatrix}
\end{equation}

\end{widetext} 
\end{document}